\documentstyle[12pt]{article}
\def\bar{\overline}

\def\1{{\chi}}

\begin{document}
\title {{The n-th root of sequential effect
algebras}\thanks{This project is supported by Natural Science
Foundation of China (10771191 and 10471124) and  Natural Science
Foundation of Zhejiang Province of China (Y6090105).}}
\author {Shen Jun$^{1,2}$, Wu Junde$^{1}$\date{}\thanks{Tel: 86-571-87951609-8111, E-mail: wjd@zju.edu.cn}}
\maketitle
$^1${\small\it Department of Mathematics, Zhejiang
University, Hangzhou 310027, P. R. China}

$^2${\small\it Department of Mathematics, Anhui Normal University,
Wuhu 241003, P. R. China}

\begin{abstract} \noindent {In 2005, Professor Gudder presented 25 open problems of sequential effect algebras, the 20th problem asked: In a sequential effect
algebra, if the square root of some element exists, is it unique ?
We can strengthen the problem as following: For each given positive
integer $n>1$, is there a sequential effect algebra such that the
n-th root of its some element $c$ is not unique and the n-th root of
$c$ is not the k-th root of $c$ ($k<n$) ? In this paper, we answer
the strengthened problem affirmatively.}
\end{abstract}

{\bf Keywords.} Effect algebra, sequential effect algebra, root.

{\bf PACS numbers:} 02.10-v, 02.30.Tb, 03.65.Ta.

\vskip0.1in

\noindent Let $H$ be a complex Hilbert space and ${\cal D}(H)$ the
set of density operators on $H$, i.e., the trace class positive
operators on $H$ of unit trace, which represent the states of
quantum system. A self-adjoint operator $A$ on $H$ such that $0\leq
A\leq I$ is called a {\it quantum effect} ([1, 2]), the set of
quantum effects on $H$ is denoted by ${\cal E}(H)$. The set of
orthogonal projection operators on $H$ is denoted by ${\cal P}(H)$.
For each $P\in {\cal P}(H)$ is associated a so-called L\"{u}ders
transformation $\Phi_L^P: {\cal D}(H)\rightarrow {\cal D}(H)$ such
that for each $T\in {\cal D}(H)$, $\Phi_L^P(T)=PTP$. Moreover, each
quantum effect $B\in {\cal E}(H)$ gives also to a general L\"{u}ders
transformation $\Phi_L^B$ such that for each $T\in {\cal D}(H)$,
$\Phi_L^B(T)=B^{\frac{1}{2}}TB^{\frac{1}{2}}$ ([3-4]).

Let $B, C\in{\cal E}(H)$ be two quantum effects. It is easy to prove
that the composition $\Phi_L^B\circ \Phi_L^C$ satisfies that for
each $T\in {\cal D}(H)$, ($\Phi_L^B\circ
\Phi_L^C)(T)=(B^{\frac{1}{2}}CB^{\frac{1}{2}})^{\frac{1}{2}}T(B^{\frac{1}{2}}CB^{\frac{1}{2}})^{\frac{1}{2}}$
([4]). Professor Gudder called $B^{\frac{1}{2}}CB^{\frac{1}{2}}$ the
{\it sequential product} of $B$ and $C$, and denoted it by $B\circ
C$ ([5-7]). This sequential product has been generalized to an
algebraic structure called a {\it sequential effect algebra} ([8]).
Now, we state the basic definitions and results of sequential effect
algebras.

\vskip 0.1 in

An {\it effect algebra} is a system $(E,0,1, \oplus)$, where 0 and 1
are distinct elements of $E$ and $\oplus$ is a partial binary
operation on $E$ satisfying that [9]:

(EA1). If $a\oplus b$ is defined, then $b\oplus a$ is defined and
$b\oplus a=a\oplus b$.

(EA2).  If $a\oplus (b\oplus c)$ is defined, then $(a\oplus b)\oplus
c$ is defined and $$(a\oplus b)\oplus c=a\oplus (b\oplus c).$$

(EA3). For each $a\in E$, there exists a unique element $b\in E$
such that $a\oplus b=1$.

(EA4). If $a\oplus 1$ is defined, then $a=0$.

\vskip 0.1 in

In an effect algebra $(E,0,1, \oplus)$, if $a\oplus b$ is defined,
we write $a\bot b$. For each $a\in (E,0,1, \oplus)$, it follows from
(EA3) that there exists a unique element $b\in E$ such that $a\oplus
b=1$, we denote $b$ by $a'$. Let $a, b\in (E,0,1, \oplus)$, if there
exists a $c\in E$ such that $a\bot c$ and $a\oplus c=b$, then we say
that $a\leq b$, if in addition, $a\neq b$, then we write $a<b$. It
follows from [9] that $\leq $ is a partial order of $(E,0,1,
\oplus)$ and satisfies that for each $a\in E$, $0\leq a\leq 1$,
$a\bot b$ if and only if $a\leq b'$.

\vskip 0.1 in

A {\it sequential effect algebra} is an effect algebra $(E,0,1,
\oplus)$ and another binary operation $\circ $ defined on $(E,0,1,
\oplus)$ satisfying that [8]:

(SEA1). The map $b\mapsto a\circ b$ is additive for each $a\in E$,
that is, if $b\bot c$, then $a\circ b\bot a\circ c$ and $a\circ
(b\oplus c)=a\circ b\oplus a\circ c$.

(SEA2). $1\circ a=a$ for each $a\in E$.

(SEA3). If $a\circ b=0$, then $a\circ b=b\circ a$.

(SEA4). If $a\circ b=b\circ a$, then $a\circ b'=b'\circ a$ and
$a\circ (b\circ c)=(a\circ b)\circ c$ for each $c\in E$.

(SEA5). If $c\circ a=a\circ c$ and $c\circ b=b\circ c$, then
$c\circ(a\circ b)=(a\circ b)\circ c$ and $c\circ(a\oplus b)=(a\oplus
b)\circ c$ whenever $a\bot b$.

\vskip 0.1 in

Let $(E,0,1, \oplus, \circ)$ be a sequential effect algebra. Then
the operation $\circ$ is said to be a {\it sequential product} on
$(E,0,1, \oplus, \circ)$. If $a, b\in (E,0,1, \oplus, \circ)$ and
$a\circ b=b\circ a$, then $a$ and $b$ is said to be {\it
sequentially independent} and write $a|b$ ([8]). Let $a\in (E,0,1,
\oplus, \circ)$. If there exists an element $b\in (E,0,1, \oplus,
\circ)$ such that $\underbrace{b\circ b\circ\cdots\circ
b}\limits_{the\ number\ is\ n}=a$, then we write $b^n=a$ and $b$ is
said to be a {\it n-th root} of $a$. Note that $b$ is a n-th root of
$a$ implies that $a$ can be obtained by measuring $b$ n-times
repeatedly.

 \vskip 0.1 in

The sequential effect algebra is an important and interesting
mathematical model for studying the quantum measurement theory
[5-8]. In [10], Professor Gudder presented 25 open problems to
motivate the study of sequential effect algebra theory. The 20th
problem asked:

\vskip 0.1 in

{\bf Problem 1} ([10]). In a sequential effect algebra $(E,0,1,
\oplus, \circ)$, if the square root of some element exists, is it
unique ?

\vskip 0.1 in

Now, we can strengthen Problem 1 as following:

\vskip 0.1 in

{\bf Problem 2}. For each given positive integer $n>1$, is there a
sequential effect algebra $(E,0,1, \oplus, \circ)$ such that the
n-th root of its some element $c$ is not unique and the n-th root of
$c$ is not the k-th root of $c$ ($k<n$) ?  i.e., are there $a,b\in
E$, such that $a\neq b$, $a^n=c=b^n$ and $a^k\neq c$, $b^k\neq c$
for $k<n$ ?

\vskip 0.1 in

In this paper, we present an example to answer Problem 2
affirmatively. Actually, we will construct a sequential effect
algebra $E_0$, such that there are elements $a,b,c\in E_0$ having
the relations
$$ a>a^2>\cdots>a^{n},$$
$$ b>b^2>\cdots>b^{n},$$
$$a^k\neq b^k\ for\ k<n\ ,\ a^{n}=b^{n}=c\neq 0.$$

\vskip 0.1 in

In order to construct our example, we need some preliminary steps:

\vskip 0.1 in

Suppose $Z$ be the integer set, $n>1$ be a given positive integer.

Let $p(x)=\sum\limits_{i=1}^{n-1}k_ix^i$, where $k_i\in Z$,
$k_i\equiv 0$ or the first nonzero $k_i>0$, we denote all the
polynomials characterized above by $I_{0}$ .

\vskip 0.1 in

Suppose $p_1,p_2\in I_{0}$ and
$p_1(x)=\sum\limits_{i=1}^{n-1}k_{1,i}x^i$ ,
$p_2(x)=\sum\limits_{i=1}^{n-1}k_{2,i}x^i$ , let
$F(p_1,p_2)(x)=\sum\limits_{i+j\leq n-1}k_{1,i}k_{2,j}x^{i+j}$ ,
$G(p_1,p_2)=\sum\limits_{i+j=n}k_{1,i}k_{2,j}$ . Then it is easy to
see that $F(p_1,p_2)\in I_{0}$ and $G(p_1,p_2)\in Z$ .

Thus we defined mappings
$$\hbox{$F:I_{0}\times I_{0}\longrightarrow I_{0}$~~~and~~~$G:I_{0}\times I_{0}\longrightarrow Z$\ .}$$

\vskip 0.1 in

Moreover, suppose $p_1,p_2,p_3\in I_{0}$ and
$p_1(x)=\sum\limits_{i=1}^{n-1}k_{1,i}x^i$ ,
$p_2(x)=\sum\limits_{i=1}^{n-1}k_{2,i}x^i$ ,
$p_3(x)=\sum\limits_{i=1}^{n-1}k_{3,i}x^i$ , let
$\bar{F}(p_1,p_2,p_3)(x)=\sum\limits_{i+j+m\leq
n-1}k_{1,i}k_{2,j}k_{3,m}x^{i+j+m}$ ,
$\bar{G}(p_1,p_2,p_3)=\sum\limits_{i+j+m=n}k_{1,i}k_{2,j}k_{3,m}$ .
Then it is also easy to see that $\bar{F}(p_1,p_2,p_3)\in I_{0}$ and
$\bar{G}(p_1,p_2,p_3)\in Z$ . Thus we defined mappings
$$\hbox{$\bar{F}:I_{0}\times I_{0}\times I_{0}\longrightarrow I_{0}$~~~and~~~$\bar{G}:I_{0}\times I_{0}\times I_{0}\longrightarrow Z$\ .}$$

\vskip 0.1 in

{\bf Lemma 1.} Suppose $p,p_1,p_2,p_3\in I_{0}$, we have

(1). $F(p_1,p_2)=F(p_2,p_1)$, $G(p_1,p_2)=G(p_2,p_1)$;

(2). $F(p_1,p_2+p_3)=F(p_1,p_2)+F(p_1,p_3)$,
$G(p_1,p_2+p_3)=G(p_1,p_2)+G(p_1,p_3)$;

(3). $F(0,p)=0$, $G(0,p)=0$;

(4). if $F(p_1,p_2)=0$, then $G(p_1,p_2)\geq 0$;

(5). $p_1-F(p_1,p_2)\in I_{0}$, and
$p_1=F(p_1,p_2)\Longleftrightarrow p_1=0$;

(6). $F(F(p_1,p_2),p_3)=\bar{F}(p_1,p_2,p_3)$,
$G(F(p_1,p_2),p_3)=\bar{G}(p_1,p_2,p_3)$;

(7). $p_1+p_2\in I_{0}$, and $p_1+p_2=0\Longleftrightarrow
p_1=p_2=0$.

{\bf Proof.} (1),(2),(3),(6) and (7) are trivial.

(4). Except for the trivial cases, we may suppose
$p_1(x)=\sum\limits_{i=n_1}^{n-1}k_{1,i}x^i$,
$p_2(x)=\sum\limits_{i=n_2}^{n-1}k_{2,i}x^i$, with $k_{1,n_1}>0$ and
$k_{2,n_2}>0$. Then from $F(p_1,p_2)=0$ we have $n_1+n_2\geq n$. If
$n_1+n_2=n$, then $G(p_1,p_2)=k_{1,n_1}k_{2,n_2}>0$; otherwise
$n_1+n_2>n$ and $G(p_1,p_2)=0$.

(5). Except for the trivial cases, we may suppose
$p_1(x)=\sum\limits_{i=n_1}^{n-1}k_{1,i}x^i$,
$p_2(x)=\sum\limits_{i=n_2}^{n-1}k_{2,i}x^i$, with $k_{1,n_1}>0$ and
$k_{2,n_2}>0$. Then the first item of $p_1-F(p_1,p_2)$ is
$k_{1,n_1}x^{n_1}$, so $p_1-F(p_1,p_2)\in I_{0}$. If $p_1\neq 0$,
then from the above reason we know that $p_1-F(p_1,p_2)\neq 0$.
Thus, the lemma is proved.

\vskip 0.1 in

Now, we take two infinite sets $U$ and $V$ such that $U\cap
V=\emptyset$. Let $f: I_{0}\times I_{0}\times Z\rightarrow U$ and
$g: I_{0}\times I_{0}\times Z\rightarrow V$ be two one to one maps.
Then, we construct our example as following:

\vskip 0.2 in

Let $E_0=\{f(p,q,m), g(p,q,m)|p,q\in I_{0}, m\in Z\ and\ satisfy\
that\ m\geq 0\ whenever\ p=q=0\}$.

\vskip 0.1 in

First, we define a partial binary operation $\oplus$ on $E_0$ as
follows (when we write $x\oplus y=z$, we always mean that $x\oplus
y=z=y\oplus x$):

(i). $f(p_1,q_1,m_1)\oplus
f(p_2,q_2,m_2)=f(p_1+p_2,q_1+q_2,m_1+m_2)$ (the right side is
well-defined, see Lemma 1(7));

(ii). for $p_2-p_1\in I_0$, $q_2-q_1\in I_0$, and satisfy that
$m_2\geq m_1$ when $p_2=p_1$ and $q_2=q_1$, $f(p_1,q_1,m_1)\oplus
g(p_2,q_2,m_2)=g(p_2-p_1,q_2-q_1,m_2-m_1)$ .

No other $\oplus$ operation is defined.

\vskip 0.1 in

Next, we define a binary operation  $\circ$  on $E_0$ as follows
(when we write $x\circ y=z$, we always mean that $x\circ y=z=y\circ
x$):

(i). $f(p_1,q_1,m_1)\circ
f(p_2,q_2,m_2)=f\Big{(}F(p_1,p_2),F(q_1,q_2),G(p_1,p_2)+G(q_1,q_2)\Big{)}$
(the right side is well-defined, see Lemma 1(4));

(ii). $f(p_1,q_1,m_1)\circ
g(p_2,q_2,m_2)=f\Big{(}p_1-F(p_1,p_2),q_1-F(q_1,q_2),m_1-G(p_1,p_2)-G(q_1,q_2)\Big{)}$
(the right side is well-defined, see Lemma 1(3), (5));

(iii). $g(p_1,q_1,m_1)\circ
g(p_2,q_2,m_2)=g\Big{(}p_1+p_2-F(p_1,p_2),q_1+q_2-F(q_1,q_2),m_1+m_2-G(p_1,p_2)-G(q_1,q_2)\Big{)}$
(the right side is well-defined, see Lemma 1(3), (5), (7)).

\vskip0.1in

We denote $f(0,0,0)$ by $0$, $g(0,0,0)$ by $1$.

\vskip0.2in

{\bf Proposition 1.} $(E_0,0,1, \oplus , \circ)$ {\it is a
sequential effect algebra}.

{\bf Proof.} In the proof below, we will use Lemma 1 frequently
without annotation. First, we verify that $(E_0,0,1, \oplus)$ is an
effect algebra.

(EA1) is obvious. We verify (EA2) as follows:

(i). $f(p_1,q_1,m_1)\oplus \Big{(}f(p_2,q_2,m_2)\oplus
f(p_3,q_3,m_3)\Big{)}=\Big{(}f(p_1,q_1,m_1)\oplus
f(p_2,q_2,m_2)\Big{)}\oplus
f(p_3,q_3,m_3)=f(p_1+p_2+p_3,q_1+q_2+q_3,m_1+m_2+m_3)$;

(ii). $f(p_1,q_1,m_1)\oplus \Big{(}f(p_2,q_2,m_2)\oplus
g(p_3,q_3,m_3)\Big{)}$ or $\Big{(}f(p_1,q_1,m_1)\oplus
f(p_2,q_2,m_2)\Big{)}\oplus g(p_3,q_3,m_3)$ is defined iff
$p_3-p_1-p_2\in I_0$, $q_3-q_1-q_2\in I_0$ and satisfy that $m_3\geq
m_1+m_2$ when $p_3=p_1+p_2$ and $q_3=q_1+q_2$, at this point, they
all equal to $g(p_3-p_1-p_2,q_3-q_1-q_2,m_3-m_1-m_2)$.

\vskip0.1in

Note that $f(p,q,m)\oplus g(p,q,m)=g(0,0,0)=1$, we verified (EA3).

For (EA4), we note from our construction that the unique element
orthogonal to $g(0,0,0)(=1)$ is $f(0,0,0)(=0)$, that is,
$f(0,0,0)\bot g(0,0,0)$ and $f(0,0,0)\oplus g(0,0,0)=g(0,0,0)$.

So far, we have proved that $(E_0,0,1, \oplus)$ is an effect
algebra.

\vskip0.1in

Next, we verify that $(E_0,0,1, \oplus , \circ)$ is a sequential
effect algebra.

(SEA3) and (SEA5) are obvious.

We verify (SEA1) as follows:

(i). $f(p_1,q_1,m_1)\circ \Big{(}f(p_2,q_2,m_2)\oplus
f(p_3,q_3,m_3)\Big{)}=f(p_1,q_1,m_1)\circ f(p_2,q_2,m_2)\oplus
f(p_1,q_1,m_1)\circ
f(p_3,q_3,m_3)=f\Big{(}F(p_1,p_2+p_3),F(q_1,q_2+q_3),G(p_1,p_2+p_3)+G(q_1,q_2+q_3)\Big{)}$,

$g(p_1,q_1,m_1)\circ \Big{(}f(p_2,q_2,m_2)\oplus
f(p_3,q_3,m_3)\Big{)}=g(p_1,q_1,m_1)\circ f(p_2,q_2,m_2)\oplus
g(p_1,q_1,m_1)\circ
f(p_3,q_3,m_3)=f\Big{(}p_2+p_3-F(p_1,p_2+p_3),q_2+q_3-F(q_1,q_2+q_3),m_2+m_3-G(p_1,p_2+p_3)-G(q_1,q_2+q_3)\Big{)}$;

(ii). when $f(p_2,q_2,m_2)\oplus g(p_3,q_3,m_3)$ is defined, i.e.,
when $p_3-p_2\in I_0$, $q_3-q_2\in I_0$, and satisfy that $m_3\geq
m_2$ if $p_3=p_2$ and $q_3=q_2$ ,

$f(p_1,q_1,m_1)\circ \Big{(}f(p_2,q_2,m_2)\oplus
g(p_3,q_3,m_3)\Big{)}=f(p_1,q_1,m_1)\circ f(p_2,q_2,m_2)\oplus
f(p_1,q_1,m_1)\circ
g(p_3,q_3,m_3)=f\Big{(}p_1-F(p_1,p_3-p_2),q_1-F(q_1,q_3-q_2),m_1-G(p_1,p_3-p_2)-G(q_1,q_3-q_2)\Big{)}$,

$g(p_1,q_1,m_1)\circ \Big{(}f(p_2,q_2,m_2)\oplus
g(p_3,q_3,m_3)\Big{)}=g(p_1,q_1,m_1)\circ f(p_2,q_2,m_2)\oplus
g(p_1,q_1,m_1)\circ
g(p_3,q_3,m_3)=g\Big{(}p_1+p_3-p_2-F(p_1,p_3-p_2),q_1+q_3-q_2-F(q_1,q_3-q_2),m_1+m_3-m_2-G(p_1,p_3-p_2)-G(q_1,q_3-q_2)\Big{)}$.

\vskip0.1in

We verify (SEA2) as follows:
$$1\circ f(p,q,m)=g(0,0,0)\circ
f(p,q,m)=f(p,q,m);$$
$$1\circ g(p,q,m)=g(0,0,0)\circ
g(p,q,m)=g(p,q,m).$$

We verify (SEA4) as follows:

\vskip0.1in

(i). $f(p_1,q_1,m_1)\circ \Big{(}f(p_2,q_2,m_2)\circ
f(p_3,q_3,m_3)\Big{)}$\\
$=f(p_1,q_1,m_1)\circ
f\Big{(}F(p_2,p_3),F(q_2,q_3),G(p_2,p_3)+G(q_2,q_3)\Big{)}$\\
$=f\Big{(}F(p_1,F(p_2,p_3)),F(q_1,F(q_2,q_3)),G(p_1,F(p_2,p_3))+G(q_1,F(q_2,q_3))\Big{)}$\\
$=f\Big{(}\bar{F}(p_1,p_2,p_3),\bar{F}(q_1,q_2,q_3),\bar{G}(p_1,p_2,p_3)+\bar{G}(q_1,q_2,q_3)\Big{)}$,

\vskip0.1in

by symmetry,

$\Big{(}f(p_1,q_1,m_1)\circ f(p_2,q_2,m_2)\Big{)}\circ
f(p_3,q_3,m_3)$\\
$=f(p_3,q_3,m_3)\circ \Big{(}f(p_1,q_1,m_1)\circ
f(p_2,q_2,m_2)\Big{)}$\\
$=f\Big{(}\bar{F}(p_1,p_2,p_3),\bar{F}(q_1,q_2,q_3),\bar{G}(p_1,p_2,p_3)+\bar{G}(q_1,q_2,q_3)\Big{)}$,

\vskip0.1in

so we have

$f(p_1,q_1,m_1)\circ \Big{(}f(p_2,q_2,m_2)\circ
f(p_3,q_3,m_3)\Big{)}=\Big{(}f(p_1,q_1,m_1)\circ
f(p_2,q_2,m_2)\Big{)}\circ f(p_3,q_3,m_3)$.

\vskip0.1in

(ii). $f(p_1,q_1,m_1)\circ \Big{(}f(p_2,q_2,m_2)\circ
g(p_3,q_3,m_3)\Big{)}$ \\
$=f(p_1,q_1,m_1)\circ
f\Big{(}p_2-F(p_2,p_3),q_2-F(q_2,q_3),m_2-G(p_2,p_3)-G(q_2,q_3)\Big{)}$ \\
$=f\Big{(}F(p_1,p_2-F(p_2,p_3)),F(q_1,q_2-F(q_2,q_3)),G(p_1,p_2-F(p_2,p_3))+G(q_1,q_2-\linebreak ~~~~~~~~~F(q_2,q_3))\Big{)}$\\
$=f\Big{(}F(p_1,p_2)-F(p_1,F(p_2,p_3)),F(q_1,q_2)-F(q_1,F(q_2,q_3)),G(p_1,p_2)-\linebreak ~~~~~~~~~~~~~~G(p_1,F(p_2,p_3))+G(q_1,q_2)-G(q_1,F(q_2,q_3))\Big{)}$ \\
$=f\Big{(}F(p_1,p_2)-\bar{F}(p_1,p_2,p_3),F(q_1,q_2)-\bar{F}(q_1,q_2,q_3),G(p_1,p_2)-\bar{G}(p_1,p_2,p_3)+\linebreak
~~~~~~~~~~G(q_1,q_2)-\bar{G}(q_1,q_2,q_3)\Big{)}$,

\vskip0.1in

$\Big{(}f(p_1,q_1,m_1)\circ f(p_2,q_2,m_2)\Big{)}\circ
g(p_3,q_3,m_3)$\\
$=f\Big{(}F(p_1,p_2),F(q_1,q_2),G(p_1,p_2)+G(q_1,q_2)\Big{)}\circ
g(p_3,q_3,m_3)$\\
$=f\Big{(}F(p_1,p_2)-F(F(p_1,p_2),p_3),F(q_1,q_2)-F(F(q_1,q_2),q_3),G(p_1,p_2)+G(q_1,q_2)-\linebreak ~~~~~~~~~~G(F(p_1,p_2),p_3)-G(F(q_1,q_2),q_3)\Big{)}$\\
$=f\Big{(}F(p_1,p_2)-\bar{F}(p_1,p_2,p_3),F(q_1,q_2)-\bar{F}(q_1,q_2,q_3),G(p_1,p_2)-\bar{G}(p_1,p_2,p_3)+\linebreak
~~~~~~~~~~G(q_1,q_2)-\bar{G}(q_1,q_2,q_3)\Big{)}$,

\vskip0.1in

so we have

$f(p_1,q_1,m_1)\circ \Big{(}f(p_2,q_2,m_2)\circ
g(p_3,q_3,m_3)\Big{)}=\Big{(}f(p_1,q_1,m_1)\circ
f(p_2,q_2,m_2)\Big{)}\circ g(p_3,q_3,m_3)$ .

\vskip0.1in

(iii). $f(p_1,q_1,m_1)\circ \Big{(}g(p_2,q_2,m_2)\circ
g(p_3,q_3,m_3)\Big{)}$ \\
$=f(p_1,q_1,m_1)\circ
g\Big{(}p_2+p_3-F(p_2,p_3),q_2+q_3-F(q_2,q_3),m_2+m_3-G(p_2,p_3)-G(q_2,q_3)\Big{)}$ \\
$=f\Big{(}p_1-F(p_1,p_2+p_3-F(p_2,p_3)),q_1-F(q_1,q_2+q_3-F(q_2,q_3)),m_1-G(p_1,p_2+\linebreak ~~~~~~~~~~p_3-F(p_2,p_3))-G(q_1,q_2+q_3-F(q_2,q_3))\Big{)}$ \\
$=f\Big{(}p_1-F(p_1,p_2+p_3)+\bar{F}(p_1,p_2,p_3),q_1-F(q_1,q_2+q_3)+\bar{F}(q_1,q_2,q_3),m_1-G(p_1,p_2+\linebreak
~~~~~~~~~~p_3)+\bar{G}(p_1,p_2,p_3)-G(q_1,q_2+q_3)+\bar{G}(q_1,q_2,q_3)\Big{)}$,

\vskip0.1in

$\Big{(}f(p_1,q_1,m_1)\circ g(p_2,q_2,m_2)\Big{)}\circ
g(p_3,q_3,m_3)$\\
$=f\Big{(}p_1-F(p_1,p_2),q_1-F(q_1,q_2),m_1-G(p_1,p_2)-G(q_1,q_2)\Big{)}\circ
g(p_3,q_3,m_3)$\\
$=f\Big{(}p_1-F(p_1,p_2)-F(p_1-F(p_1,p_2),p_3),q_1-F(q_1,q_2)-F(q_1-F(q_1,q_2),q_3),m_1-\linebreak ~~~~~~~~~~G(p_1,p_2)-G(q_1,q_2)-G(p_1-F(p_1,p_2),p_3)-G(q_1-F(q_1,q_2),q_3)\Big{)}$\\
$=f\Big{(}p_1-F(p_1,p_2+p_3)+\bar{F}(p_1,p_2,p_3),q_1-F(q_1,q_2+q_3)+\bar{F}(q_1,q_2,q_3),m_1-G(p_1,p_2+\linebreak
~~~~~~~~~~p_3)+\bar{G}(p_1,p_2,p_3)-G(q_1,q_2+q_3)+\bar{G}(q_1,q_2,q_3)\Big{)}$,

\vskip0.1in

so we have

$f(p_1,q_1,m_1)\circ \Big{(}g(p_2,q_2,m_2)\circ
g(p_3,q_3,m_3)\Big{)}=\Big{(}f(p_1,q_1,m_1)\circ
g(p_2,q_2,m_2)\Big{)}\circ g(p_3,q_3,m_3)$.

\vskip0.1in

(iv). $g(p_1,q_1,m_1)\circ \Big{(}g(p_2,q_2,m_2)\circ
g(p_3,q_3,m_3)\Big{)}$ \\
$=g(p_1,q_1,m_1)\circ
g\Big{(}p_2+p_3-F(p_2,p_3),q_2+q_3-F(q_2,q_3),m_2+m_3-G(p_2,p_3)-G(q_2,q_3)\Big{)}$ \\
$=g\Big{(}p_1+p_2+p_3-F(p_2,p_3)-F(p_1,p_2+p_3-F(p_2,p_3)),q_1+q_2+q_3-F(q_2,q_3)-\linebreak ~~~~~~~~~~F(q_1,q_2+q_3-F(q_2,q_3)),m_1+m_2+m_3-G(p_2,p_3)-G(q_2,q_3)-G(p_1,p_2+\linebreak ~~~~~~~~~~p_3-F(p_2,p_3))-G(q_1,q_2+q_3-F(q_2,q_3))\Big{)}$ \\
$=g\Big{(}p_1+p_2+p_3-F(p_2,p_3)-F(p_1,p_2)-F(p_1,p_3)+\bar{F}(p_1,p_2,p_3),q_1+q_2+q_3-\linebreak
~~~~~~~~~~F(q_2,q_3)-F(q_1,q_2)-F(q_1,q_3)+\bar{F}(q_1,q_2,q_3),m_1+m_2+m_3-G(p_2,p_3)-\linebreak
~~~~~~~~~~G(p_1,p_2)-G(p_1,p_3)+\bar{G}(p_1,p_2,p_3)-G(q_2,q_3)-G(q_1,q_2)-G(q_1,q_3)+\linebreak
~~~~~~~~~~~~~~\bar{G}(q_1,q_2,q_3)\Big{)}$,

\vskip0.1in

by symmetry, we have

$g(p_1,q_1,m_1)\circ \Big{(}g(p_2,q_2,m_2)\circ
g(p_3,q_3,m_3)\Big{)}=\Big{(}g(p_1,q_1,m_1)\circ
g(p_2,q_2,m_2)\Big{)}\circ g(p_3,q_3,m_3)$.

\vskip0.1in

Thus, we proved that $(E_0,0,1, \oplus , \circ)$ is a sequential
effect algebra and the theorem is proved.

\vskip0.3in

Now, let $P_i(x)=x^i$. Then it is easy to see that
$$
F(P_1,P_j)= \left\{
  \begin{array}{ll}
    P_{1+j}\ , & \hbox{$if\ j<n-1$;} \\
    0\ , & \hbox{$if\ j=n-1$.}
  \end{array}
\right.~and~~G(P_1,P_j)= \left\{
  \begin{array}{ll}
    0\ , & \hbox{$if\ j<n-1$;} \\
    1\ , & \hbox{$if\ j=n-1$.}
  \end{array}
\right.
$$ Thus we have
$$\hbox{$[f(P_1,0,0)]^{k}=f(P_1,0,0)\circ f(P_{k-1},0,0)=f(P_{k},0,0)$ for $k<n$,}$$
$$\hbox{$[f(P_1,0,0)]^{n}=f(P_1,0,0)\circ f(P_{n-1},0,0)=f(0,0,1)$,}$$
$$\hbox{$[f(P_1,0,0)]^{n+1}=f(P_1,0,0)\circ f(0,0,1)=0$,}$$ and
$$\hbox{$[f(0,P_1,0)]^{k}=f(0,P_1,0)\circ f(0,P_{k-1},0)=f(0,P_{k},0)$ for $k<n$,}$$
$$\hbox{$[f(0,P_1,0)]^{n}=f(0,P_1,0)\circ f(0,P_{n-1},0)=f(0,0,1)$,}$$
$$\hbox{$[f(0,P_1,0)]^{n+1}=f(0,P_1,0)\circ f(0,0,1)=0$.}$$

If we denote $f(P_1,0,0)$ by $a$, $f(0,P_1,0)$ by $b$, $f(0,0,1)$ by
$c$, then it is easy to get the relations
$$ a>a^2>\cdots>a^{n}>a^{n+1},$$
$$ b>b^2>\cdots>b^{n}>b^{n+1},$$
$$a^k\neq b^k\ for\ k<n\ ,\ a^{n}=b^{n}=c\neq 0\ and\ a^{n+1}=b^{n+1}=0
.$$ That is, $a, b$ are the n-th root of $c$, but $a, b$ are not the
k-th root of $c$, where $k=2, 3, \cdots, n-1$, moreover, $a, b$ are
also the n+1-th root of $0$, so, the Problem 2 is answered
affirmatively.

\vskip 0.1 in

Finally, we would like to point out that for the advances of
sequential effect algebras, see [11-16].

\vskip 0.2 in

\centerline {\bf Acknowledgement}

\vskip 0.2 in

The authors wish to express their thanks to the referee for his
valuable comments and suggestions.

\vskip 0.2 in

\vskip0.2in

\centerline{\bf References}

\vskip0.2in

\noindent [1]. Ludwig, G. {\it Foundations of Quantum Mechanics
(I-II)}, Springer, New York, 1983.

\noindent [2]. Ludwig, G. {\it An Axiomatic Basis for Quantum
Mechanics (II)}, Springer, New York, 1986.

\noindent [3]. Davies, E. B. {\it Quantum Theory of Open Systems},
Academic Press, London, 1976.

\noindent [4]. Busch, P, Grabowski, M and Lahti P. J, {\it
Operational Quantum Physics}, Springer-Verlag, Beijing Word
Publishing Corporation, 1999.

\noindent [5]. Gudder, S, Nagy, G. Sequential quantum measurements.
J. Math. Phys. 42(2001), 5212-5222.

\noindent [6]. Gheondea, A, Gudder, S. Sequential product of quantum
effects. Proc. Amer. Math. Soc. 132 (2004), 503-512.

\noindent [7]. Gudder, S, Latr¨¦moli¨¨re, F. Characterization of the
sequential product on quantum effects. J. Math. Phys. 49 (2008),
052106-052112.

\noindent [8]. Gudder, S, Greechie, R. Sequential products on effect
algebras. Rep. Math. Phys.  49(2002), 87-111.

\noindent [9]. Foulis, D J, Bennett, M K. Effect algebras and
unsharp quantum logics. Found Phys 24 (1994), 1331-1352.

\noindent [10]. Gudder, S. Open problems for sequential effect
algebras. Inter. J. Theory. Physi. 44 (2005), 2219-2230.

\noindent [11] Shen Jun and Wu Junde. Not each sequential effect
algebra is sharply dominating. Phys. Letter A. {\bf 373}, 1708-1712,
(2009)

\noindent [12] Shen Jun and Wu Junde. Remarks on the sequential
effect algebras. Report. Math. Phys. {\bf 63}, 441-446, (2009)

\noindent [13] Shen Jun and Wu Junde. Sequential product on standard
effect algebra ${\cal E}(H)$. J. Phys. A: Math. Theor. {\bf 44},
345203-345214, (2009)

\noindent [14] Shen Jun and Wu Junde. The Average Value Inequality
in Sequential Effect Algebras. Acta Math. Sinica, English Series.
{\bf 25}, 1330-1336, (2009)

\noindent [15] Liu Weihua and Wu Junde. A uniqueness problem of the
sequence product on operator effect algebra ${\cal E}(H)$. J. Phys.
A: Math. Theor. {\bf 42}, 185206-185215, (2009)

\noindent [16] Liu Weihua and Wu Junde. On fixed points of
L\"{u}ders operation. J. Math. Phys. {\bf 50}, 103531-103532, (2009)

\end{document}